\documentclass[twocolumn,aps,floatfix,superscriptaddress,showpacs]{revtex4}
\usepackage{amsmath,amssymb,eucal,graphicx}
\begin{document}
\title{First Passage Properties of the P\'olya Urn Process}
\author{Tibor Antal}
\affiliation{Program for Evolutionary Dynamics,
  Harvard University, Cambridge, MA 02138, USA}
\author{E.~Ben-Naim}
\affiliation{Theoretical Division and Center for Nonlinear
Studies, Los Alamos National Laboratory, Los Alamos, New Mexico
87545, USA}
\author{P.~L.~Krapivsky}
\affiliation{Department of Physics,
Boston University, Boston, Massachusetts 02215, USA}
\begin{abstract}
We study first passage statistics of the P\'olya urn model.  In this
random process, the urn contains two types of balls. In each step, one
ball is drawn randomly from the urn, and subsequently placed back into
the urn together with an additional ball of the same type. We derive
the probability $G_n$ that the two types of balls are equal in number,
for the first time, when there is a total of $2n$ balls.  This first
passage probability decays algebraically, $G_n\sim n^{-2}$, when $n$
is large. We also derive the probability that a tie ever happens. This
probability is between zero and one, so that a tie may occur in some
realizations but not in others.  The likelihood of a tie is appreciable
only if the initial difference in the number balls is of the order of
the square-root of the total number of balls.
\end{abstract}
\maketitle

\section{Introduction}

Urn models play a central role in probability theory and 
combinatorics \cite{wf,msb}.  Since the balls can represent anything
from atoms to biological organisms to humans, urn models are widely
used in the physical, life, and social sciences \cite{jk}.

In this paper, we investigate the classic P\'olya urn model
\cite{aam,ep,hmm}. This urn process is a type of birth process, and it
is useful for modeling the spread of infectious diseases, population
dynamics, and evolutionary processes in biology
\cite{ep,ks,fh,bp,kkkr}. Furthermore, this stochastic process is a
branching process \cite{teh}, and it is used to model data
structures in computer science \cite{bp1,afp,kmr}. From the myriad of
other applications, we mention decision making \cite{mh},
reinforcement learning \cite{er}, and Internet browser usage
\cite{pw}. We also note that the P\'olya urn model is a limiting case
of earlier urn models investigated by Laplace \cite{psl}, Markov
\cite{aam1}, and Ehrenfest \cite{ee}.

The P\'olya urn model exhibits rich and interesting phenomenology that
includes strong influence of the initial conditions, large
realization-to-realization fluctuations, and substantial finite-size 
corrections \cite{hmm,fcp,bk}. In this study, we obtain the first
passage properties \cite{sr} of the P\'olya urn model, and contrast
these with the first passage characteristics of an ordinary random
walk \cite{sr,rg}.

In the P\'olya urn model, there are two types of balls, black and
white. In a basic step, one ball is selected randomly from all balls
in the urn. This ball is then returned to the urn together with an
additional ball of the same color. Starting with a given 
configuration of balls, the number of balls increases indefinitely by
repeating this step ad infinitum. Thus, a configuration $(B,W)$ with
$B$ black balls and $W$ white balls evolves according to
\begin{equation}
\label{process}
(B,W)\to
\begin{cases}
(B+1,W) & {\rm with\ probability}\quad   \tfrac{B}{B+W},\\
(B,W+1) & {\rm with\ probability}\quad   \tfrac{W}{B+W}.
\end{cases}
\end{equation}

\begin{figure}[t]
\includegraphics[width=0.25\textwidth]{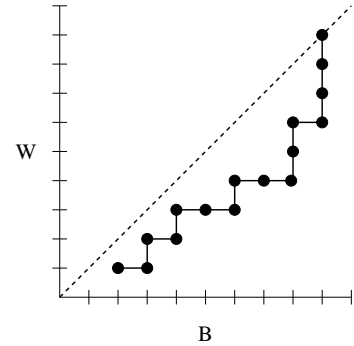}
\caption{The urn process as a trajectory on a two-dimensional lattice
(bold line) where bullets indicate intermediate stages of the
trajectory.  Exit from the $B>W$ region is equivalent to this
trajectory reaching the diagonal (broken line).}
\label{fig-urn}
\end{figure}

We investigate first passage properties of the urn process (figure
\ref{fig-urn}).  We find that the probability $G_n$ that a tie is
reached, for the first time, when there are $n$ balls of each type
decays algebraically
\begin{equation}
G_n\sim n^{-2},
\end{equation}
for large $n$. This asymptotic behavior holds for arbitrary initial
conditions. We also show that the total exit probability, that is, the
probability that a tie is ever reached, is less than one. Hence, an
initial imbalance in the number of balls can be locked in forever. We
study how the total exit probability depends on the initial condition
and find that it is appreciable only when the imbalance in the number
of balls is of the order of square-root of the total number of balls.

The rest of this paper is organized as follows. We derive the first
passage probability in section II.  We then obtain the exit
probability by summing the first passage probability (section III). We
discuss the extreme cases of nearly-maximal and extremely small exit
probabilities (sec.~IV), and then use these limiting behaviors to
establish scaling properties of the exit probability (sec.~V). We
generalize the results to near-ties in section VI, and conclude in
section VII.

\section{The first passage probability}
\label{first}

Our goal is to quantify the first passage process, illustrated in
figure \ref{fig-urn}, using the first passage probability and the
total exit probability.  For the initial condition
\hbox{$(B,W)=(b,w)$} where, without loss of generality, black balls
are in the majority, $b>w$, the first passage probability $G_n(b,w)$
is the likelihood that a tie is reached, for the first time, when
$(B,W)=(n,n)$.  In other words $G_n(b,w)$ is the probability that the
initial imbalance holds, $B>W$, if and only if $W<n$.  The total exit
probability $E(b,w)$ is the probability that a tie is ever reached.

These first passage characteristics are of interest in a variety of
contexts. For example, in growth of bacterial colonies \cite{qz,ohrn},
when bacteria proliferate without resource limitations, the exit
probability measures the likelihood that the minority species
eventually overtakes the majority species. In the context of a
branching process, first passage statistics quantify the likelihood
that two branches of a tree reach perfect balance.

As a preliminary step to finding the first passage probability, we
obtain the likelihood that the system reaches configuration $(B,W)=(m,
n)$ starting from $(B,W)=(b,w)$.  Let's consider, for example, the
transition $(1,1)\to (3,3)$ where one possible path is
\begin{equation*}
(1,1)\to(1,2)\to(1,3)\to(2,3)\to(3,3)\,.
\end{equation*}
The likelihood of this path is
\begin{equation}
\label{example}
\frac{1}{2}\times\frac{2}{3}\times\frac{1}{4}\times\frac{2}{5}=\frac{(1\cdot
2)\cdot(1\cdot 2)}{2\cdot 3\cdot 4\cdot 5}\,.
\end{equation}
There are $\binom{4}{2}=6$ distinct routes from $(1,1)$ to $(3,3)$ and
they all have the same same probability \eqref{example}.

In general, all paths from configuration $(b,w)$ to configuration
$(m,n)$ have the same probability
\begin{equation*}
\frac{[b(b+1)\cdots (m-1)]\cdot[w(w+1)\cdots
(n-1)]}{(b+w)(b+w+1)\cdots(m+n-1)}\,.
\end{equation*}
We rewrite this probability using factorials
\begin{equation*}
\frac{(m-1)!}{(b-1)!}\times\frac{(n-1)!}{(w-1)!}
\times\frac{(b+w-1)!}{(m+n-1)!}\,.
\end{equation*}
The total number of distinct paths from $(b,w)$ to $(m,n)$ equals the
binomial $\binom{m+n-b-w}{m-b}$. Hence, the transition probability $P$
that, starting from configuration $(b,w)$, the system reaches
configuration $(m,n)$ is \cite{jk,aam,ep,hmm}
\begin{equation}
\label{urn}
 P = \binom{m-1}{b-1} \binom{n-1}{w-1} \binom{m+n-1}{b+w-1}^{-1}.
\end{equation}
In particular, the probability distribution is flat \cite{aam,ep}
$P=\frac{1}{m+n-1}$, for the initial condition $(b,w)=(1,1)$.

The number of paths from $(b,w)$ to $(n,n)$ that reach the diagonal
$B=W$ only at the end point equals $\frac{b-w}{2n-b-w}$ times the
total number of such paths.  This result can be established using the
reflection principle \cite{wf}. Since all paths from $(b,w)$ to
$(n,n)$ are equiprobable, the first passage probability is simply
\hbox{$G_n(b,w)=\frac{b-w}{2n-b-w}P$}. By substituting $m=n$ into
Eq.~\eqref{urn}, we obtain our first main result, the first passage
probability
\begin{equation}
\label{G-sol}
G_n(b,w) = \frac{b-w}{b+w} \binom{n-1}{b-1}
\binom{n-1}{w-1} \binom{2n-1}{b+w}^{-1}.
\end{equation}
This quantity decays algebraically,
\begin{equation}
\label{asym}
G_n(b,w)\simeq A(b,w)\,n^{-2},
\end{equation}
in the asymptotic limit $n\gg b,w$. The proportionality constant in
\eqref{asym} is
\begin{equation*}
A(b,w)=\frac{(b-w)(b+w-1)!}{(b-1)! (w-1)!}\,2^{-b-w}\,.
\end{equation*}

For the special case $(b,w)=(2,1)$, the first passage probability
\eqref{G-sol} is simply
\begin{equation}
\label{Gn-21}
G_n(2,1)=\frac{1}{(2n-3)(2n-1)}.
\end{equation}
The probability that a tie is ever reached equals the sum
$\frac{1}{1\cdot 3}+\frac{1}{3\cdot 5}+\frac{1}{5\cdot
7}+\cdots=\frac{1}{2}$, and hence, there is a finite chance that the
initial imbalance in the number of balls is maintained forever. This
behavior is different than that of a one-dimensional random walk.  In
an ordinary random walk, the two elementary transitions in
\eqref{process} occur with probability $1/2$, and the first passage
probability decays algebraically, $G_n\sim n^{-3/2}$, for large
$n$. Yet, the exit probability equals one, and the random walk is
guaranteed to reach the diagonal $B=W$. Thus, the one-dimensional
random walk is recurrent, but the P\'olya urn process is transient.

\section{The exit probability}
\label{exit}

The exit probability ${\cal E}_n(b,w)$ is the likelihood that starting
from configuration $(b,w)$, a tie happens by the time the urn contains
$2n$ balls. The exit probability follows from the first passage
probability,
\begin{equation}
\label{En-sum}
{\cal E}_n(b,w)=\sum_{b\leq j\leq n}G_j(b,w)\,.
\end{equation}
The lower limit reflects that the quickest tie occurs when
$(B,W)=(b,b)$.  We are especially interested in the total exit
probability, $E(b,w)\equiv \lim_{n\to\infty}{\cal E}_n(b,w)$.  From
the identity ${\cal E}_{n}-{\cal E}_{n-1}(b,w)=G_n(b,w)$ and equation
\eqref{asym}, we conclude the asymptotic behavior
\begin{equation}
\label{asym-E}
E(b,w)-{\cal E}_n(b,w)\simeq  A(b,w)\,n^{-1},
\end{equation}
when $n\gg b,w$. In particular, the quantity \hbox{${\cal
E}_n(2,1)=\frac{n-1}{2n-1}$} that is the sum of \eqref{Gn-21}, agrees
with \eqref{asym-E}.

To evaluate the total exit probability $E(b,w)$, we introduce the
shorthand notation $C_k(b,w)\equiv G_{b+k}(b,w)$. With this notation,
Eq.~\eqref{En-sum} becomes \hbox{$E(b,w)=\sum_{k\geq 0}C_k(b,w)$}, and
\begin{equation}
C_k (b,w)=  \frac{b-w}{b+w}
\frac{\displaystyle \binom{b+k-1}{b-1} \binom{b+k-1}{w-1}}
{\displaystyle \binom{2b+2k-1}{b+w}},
\end{equation}
which is obtained by substituting $n=b+k$ into \eqref{G-sol}.  In
particular, the quantity
\begin{equation*}
C_0(b,w)=\frac{\Gamma(b)\,\Gamma(b+w)}{\Gamma(2b)\,\Gamma(w)}
\end{equation*}
is the probability that a tie occurs as quickly as possible.

In terms of the quantities $C_k(b,w)$, the total exit probability
$E(b,w)=\sum_{j\geq b}G_j(b,w)$ equals
\begin{equation}
\label{sum2} E(b,w)=\sum_{k\ge0}C_k(b,w).
\end{equation}
We now evaluate the ratio of two consecutive first passage
probabilities
\begin{equation}
\label{ratio}
\frac{C_{k+1}(b,w)}{C_k(b,w)} =
\frac{(k+b)(k+\frac{b-w}{2})(k+\frac{b-w+1}{2})}
{(k+1)(k+b+\frac{1}{2})(k+b-w+1)}.
\end{equation}
Given these ratios, the exit probability can be expressed in terms of
the hypergeometric function \cite{as}
\begin{equation}
\label{gensol}
E(b,w) =
\tfrac{\Gamma(b)\Gamma(b+w)}{\Gamma(2b)\Gamma(w)}
F(b,\tfrac{b-w}{2},\tfrac{b-w+1}{2};b\!+\!\tfrac{1}{2},b\!-\!w\!+\!1;1).
\end{equation}
This closed form expression is our second main result. As expected,
$E(b,b)=1$ and $E(b,0)=0$.

We also note that the exit probability satisfies the compact
recursion relation
\begin{equation}
\label{recursion} E(b,w) =
\frac{b}{b+w}\,E(b+1,w)+\frac{w}{b+w}\,E(b,w+1),
\end{equation}
for all $b\neq w$. The boundary conditions are $E(b,b)=1$ and $E(b,0)=0$.
This recursion follows directly from the definition of
the stochastic process \eqref{process}, and is reminiscent of the
recursion equation for an ordinary random walk \hbox{$E(b,w) =
\frac{1}{2}\,E(b+1,w)+\frac{1}{2}\,E(b,w+1)$} \cite{sr}. We use this recursion
to analyze extremal properties of $E(b,w)$ in the next section.

\section{Extremal Behavior}
\label{extremal}

Intuitively, we expect that when $b$ is fixed, the exit probability
increases monotonically with $w$. The exit probability is largest when the
number of balls is balanced, $E(b,b)=1$, and conversely, the exit
probability is smallest when the initial imbalance is maximal,
$E(b,0)=0$. We now discuss the extreme cases of very small and
nearly-maximal exit probabilities, respectively.

When $w=1$, the exit probability decays exponentially with the total
number of balls,
\begin{equation}
\label{E1}
 E(b,1)=2^{1-b}.
\end{equation}
To obtain this result, we note that when $w=1$, two of the
arguments of the hypergeometric function in \eqref{gensol} coincide
and hence,
\hbox{$E(b,1) = \tfrac{\Gamma(b)\,\Gamma(b+1)}{\Gamma(2b)}\,
F\left(\tfrac{b}{2}, \tfrac{b-1}{2}; b+\tfrac{1}{2};1\right)$}.
We obtain the expression \eqref{E1} using the Gauss identity for the
hypergeometric function
\begin{equation}
\label{Gauss}
F(x,y;z;1)=\frac{\Gamma(z-x-y)\,\Gamma(z)}{\Gamma(z-x)\,\Gamma(z-y)},
\end{equation}
and the following two identities for the Gamma function,
\hbox{$\Gamma(x+1)=x\Gamma(x)$}, and \hbox{$\Gamma(\frac{1}{2})\Gamma(2x)=
2^{2x-1}\Gamma(x)\Gamma\left(x+\frac{1}{2}\right)$}.

By substituting $E(b,1)=2^{1-b}$ into the recursion
\eqref{recursion}, we have $E(b,2)=(b+2)2^{-b}$. Similarly, we obtain
$E(b,3)=(b^2+5b+8)2^{-b-2}$ by substituting $E(b,2)$ into
\eqref{recursion}. In general, the exit probability has the form
\begin{equation}
\label{poly} E(b,w)=\frac{U_w(b)}{(w-1)!}\,2^{2-w-b},
\end{equation}
where $U_w(b)$ is a polynomial of degree $w-1$ in the variable $b$.
From equation \eqref{recursion}, these polynomials satisfy the
recursion
\begin{equation}
\label{rec2} U_{w+1}(b)=2(b+w)U_w(b)-b\,U_w(b+1).
\end{equation}
Starting with the boundary condition, $U_1(b)=1$, we have
\begin{equation}
\label{uwb} U_w(b)=
\begin{cases}
1&w=1,\\
b+2&w=2,\\
b^2+5b+8&w=3, \\
b^3+9b^2+32b+48&w=4,\\
b^4 + 14b^3  + 83b^2  + 262b + 384&w=5.
\end{cases}
\end{equation}
Since the coefficient of the dominant term in $U_w(b)$ equals one, the
exit probability decays exponentially with the total initial
population
\begin{equation}
\label{asy} E(b,w) \simeq \frac{b^{w-1}}{(w-1)!}2^{2-b-w},
\end{equation}
when $w$ is finite and $b\to \infty$.

To analyze the behavior in the opposite limit of nearly-maximal exit
probabilities, we consider the special case $w=b-1$ where the ratio
\eqref{ratio} simplifies as follows
\begin{equation}
\label{ratio1more}
 \frac{C_{k+1}}{C_k} = \frac{(k+b)(k+\frac{1}{2})}
 {(k+2)(k+b+\frac{1}{2})}.
\end{equation}
We now shift the index of the first passage probability by one,
$D_{k+1}\equiv C_k$, with $D_0=-1$, and then evaluate the sum
\eqref{sum2} to find $E(b,b-1) =1-
F\left(b-1,-\tfrac{1}{2};b-\tfrac{1}{2}; 1\right)$.  Further, we
express the exit probability through Gamma functions by using the
identity \eqref{Gauss},
\begin{equation}
\label{escdzero} E(b,b-1) =
1-\frac{\Gamma(b-1/2)}{\Gamma(b)\Gamma(1/2)}.
\end{equation}
The exit probability increases monotonically with $b$:
$E(b,b-1)=1/2,5/8,11/16$ for $b=2,3,4$. Moreover, ties become
practically certain, $E(b,b-1)\simeq 1-1/\sqrt{\pi b}$, in the limit
$b\to\infty$.

\begin{figure}[t]
\includegraphics[width=0.45\textwidth]{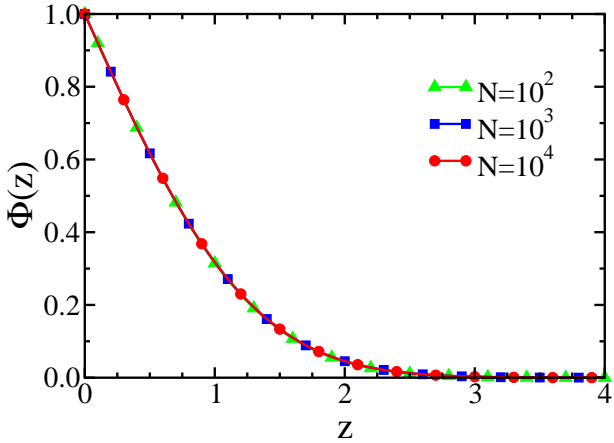}
\caption{The scaling function $\Phi(z)$ versus the scaling variable
$z$. Shown is the exit probability $E(b,w)$, evaluated using the
exact expression \eqref{gensol}, for three different values of
$N=b+w$.}
\label{fig-scaling}
\end{figure}

Along the same lines, we evaluate $E(b,b-q)$ by substituting the form
\eqref{escdzero} and the boundary condition $E(b,b)=1$ into the
recursion \eqref{recursion},
\begin{equation}
\label{escdiffq}
E(b,b-q)=1-\frac{\Gamma(b-1/2)}{\Gamma(b)\Gamma(1/2)}\times
\begin{cases}
1&q=1,\\
 2&q=2, \\
3\frac{b-5/3}{b-3/2}&q=3,\\
4\frac{b-2}{b-3/2}&q=4.
\end{cases}
\end{equation}
From these examples, we conclude $E(b,b-q)\simeq 1-q/\sqrt{\pi b}$
when $q$ is finite and $b\to\infty$. In other words,
\begin{equation}
\label{scaling-like}
 E(b,w) \simeq 1-\sqrt{\frac{2}{\pi}}\frac{|b-w|}{\sqrt{b+w}},
\end{equation}
when the initial imbalance $|b-w|$ is fixed and the total number of
balls $b+w$ diverges. We used the symmetry $E(x,y)=E(y,x)$ so that
\eqref{scaling-like} applies for both $b>w$ and $w>b$. This equation
implies that a tie is nearly certain whenever the initial discrepancy
in the number of balls is much smaller than the square-root of the
total number of balls. Otherwise, the exit probability is
substantially reduced.

\section{Typical Behavior}
\label{typical}

The asymptotic behavior \eqref{scaling-like} suggests that the exit
probability is function of a single variable when the total number of
balls is very large. Specifically, the scaling function $\Phi(z)$, given by
\begin{equation}
\label{scaling}
E(b,w) \simeq \Phi(z)\quad {\rm with}\quad z=
\frac{|b-w|}{\sqrt{b+w}},
\end{equation}
quantifies the exit probability in the limit $b+w\to\infty$. Numerical
evaluation of the exit probability \eqref{gensol} confirms this
scaling behavior (figure \ref{fig-scaling}).

\begin{figure}[t]
\includegraphics[width=0.45\textwidth]{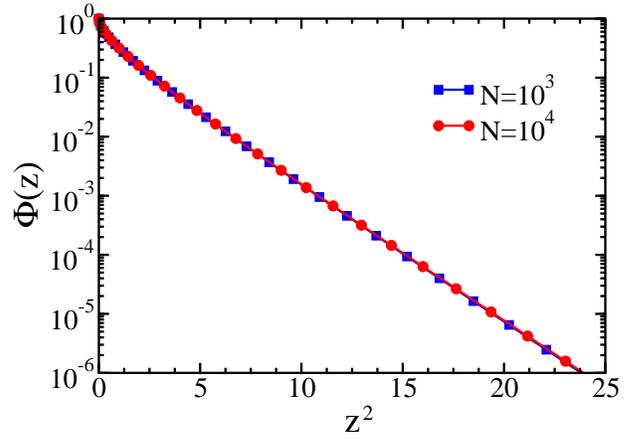}
\caption{The tail of the scaling function $\Phi(z)$. Shown is the
scaling function  $\Phi(z)$ versus $z^2$.}
\label{fig-tail}
\end{figure}

The scaling function is monotonically decreasing because a larger
initial imbalance implies a smaller exit probability. The small
argument behavior of the scaling function, $\Phi(z)\simeq
1-\sqrt{\frac{2}{\pi}}\,z$, follows immediately from
\eqref{scaling-like}. The large argument behavior follows from the
exponential behavior $E(b,1)\sim 2^{-b}$, see \eqref{E1}. Indeed,
when $w=1$ and $b$ is very large, the scaling variable is
\hbox{$z\simeq \sqrt{b}$}, and hence the tail of the scaling function
must be Gaussian, \hbox{$\log \Phi(z)\sim - z^2$}.  Numerical
evaluation of the exact solution supports this heuristic prediction
(see figure \ref{fig-tail}).

The scaling form \eqref{scaling} implies that the likelihood of a tie
is appreciable only when the initial population difference,
$\Delta=|b-w|$, is of the same order as the square-root of the total
population, $N=b+w$, that is,
\begin{equation}
\Delta \sim \sqrt{N}.
\end{equation}
A tie is nearly certain when the discrepancy is small, $\Delta\ll \sqrt{N}$,
but extremely rare when $\Delta\gg \sqrt{N}$.

\section{Near ties}
\label{gen}

There are a number of generalizations of the first passage process
discussed above. Two natural questions are: (i) what is the
probability that the {\em ratio} between the majority population and
the majority population is always above a fixed threshold and (ii)
what is the probability that the {\em difference} between the two
populations is always above a fixed threshold. In this section, we
address the latter problem.

We define $G_n(b,w;d)$ to be the first passage probability that
starting with configuration $(b,w)$, the difference \hbox{$B-W>d$} if
and only if $W<n$. In other words $G_n(b,w;d)$ is the probability that
there are at least $d$ more black balls throughout the evolution and
moreover, this condition is violated for the first time when
$(B,W)=(n+d,n)$. We obtain the first passage probability
\begin{equation}
\label{Gn_bwd}
 G_n(b,w;d) = \frac{b-w-d}{b+w} 
\frac{\displaystyle \binom{n+d-1}{b-1} \binom{n-1}{w-1}}
 {\displaystyle \binom{2n+d-1}{b+w}},
\end{equation}
by multiplying the probability \eqref{urn} for transitioning from
$(b,w)$ to $(n+d,n)$ with the fraction $\frac{b-w-d}{2n+d-b-w}$ of
these paths that do not cross the line $B=W+d$ \cite{wf}.  Of course,
this expression matches \eqref{G-sol} when $d=0$. Again, the
asymptotic behavior in the limit $n\to\infty$ is $G_n\sim n^{-2}$.

The exit probability $E(b,w;d)$ is the probability that the line
$B=W+d$ is reached at least once during the evolution. By repeating
the steps leading to \eqref{gensol}, we find the exit probability in
terms of a higher-order hypergeometric function
\begin{equation}
\label{hyper-43}
E(b,w;d)=
\tfrac{\Gamma(b-d)\,\Gamma(b+w)}{\Gamma(2b-d)\,\Gamma(w)}\,
F(c_1,c_2,c_3,c_4;e_1,e_2,e_3;1).
\end{equation}
The corresponding arguments are
\begin{equation*}
\begin{split}
c_1=b, \quad c_2=\tfrac{b-w-d}{2}, \quad c_3=\tfrac{b-w-d+1}{2}, 
\quad c_4=b-d,\\
e_1=b+\tfrac{1-d}{2}, \quad e_2=b-w-d+1, \quad e_3=b-\tfrac{d}{2}.
\end{split}
\end{equation*}

For near-ties, $d=1$, there are many similarities with $E(b,w)$ of
Eq.~\eqref{gensol}.  For example, the exit probability decays
exponentially when the initial imbalance is maximal,
\begin{equation}
E(b,1;1)=\frac{b}{b-1}2^{1-b}.
\end{equation}
Additionally, the exit probability is close to one when the initial
imbalance is minimal,
\begin{equation}
E(b,b-2;1)=1-\frac{1}{2(b-1)} - \frac{\Gamma(b-1/2)}{\Gamma(b)\Gamma(1/2)}.
\end{equation}
Thus, the behavior $E(b,b-2;1)\simeq 1-1/\sqrt{\pi\,b}$ is recovered
when $b\to\infty$.

\section{Discussion}

In summary, we obtained first passage characteristics of the P\'olya
urn process as a function of the initial condition.  The first passage
probability that a tie is reached for the first time when there are
$2n$ balls decays algebraically, $G_n\sim n^{-2}$, for large $n$.  The
probability that a tie ever occurs, is less than one, hence ties are
not certain.  This exit probability decreases as the initial
discrepancy in the number of balls increases. Moreover, there is a
universal scaling behavior when the total initial population is very
large. This scaling behavior implies that the exit probability is
appreciable only when the initial population imbalance is of the order
of the square-root of the total population.

The key property of the P\'olya urn model is that the fraction of
white balls approaches a limiting value, but this value fluctuates
from realization-to-realization. In many other urn models, however,
the opposite is true, and moreover, the two fractions approach the
same limiting value. This is the case for the Friedman urn process
\cite{fm,ff} which, in its simplest form, is equivalent to the
stochastic process
\begin{equation*}
(B,W)\to
\begin{cases}
(B+1,W) & {\rm with\ probability}\quad   \tfrac{W}{B+W},\\
(B,W+1) & {\rm with\ probability}\quad   \tfrac{B}{B+W}.
\end{cases}
\end{equation*}
It will be interesting to investigate first passage properties of this
urn process. We conjecture that first passage statistics are much
closer to those of the ordinary random walk.

In the context of population dynamics and evolutionary biology,
continuous time processes are more appropriate. The continuous time
analog of the P\'olya urn model \eqref{process} is the two-species
branching process, $B\to B+B$, and $W\to W+W$, where the two birth
rates are equal. This continuous time process is closely related to
the discrete time urn process. For instance, the total exit
probabilities for the two processes are {\em identical}.

\acknowledgments{We gratefully acknowledge support from the John
Templeton Foundation, NSF/NIH grant R01GM078986, DOE grant
DE-AC52-06NA25396, and NSF grant CCF-0829541.}

\end{document}